# Optimized Flooding Protocol for Ad hoc Networks

Vamsi K Paruchuri[a], Arjan Durresi[b], Raj Jain[b]
[a]Department of Electrical Engineering
[b]Department of Computer and Information Science
The Ohio State University
2015 Neil Avenue, Columbus, OH 43221
paruchuri.1@osu.edu, {durresi, jain}@cis.ohio-state.edu

***Abstract*— Flooding provides important control and route establishment functionality for a number of unicast and multicast protocols in Mobile Ad Hoc Networks. Considering its wide use as a building block for other network layer protocols, the flooding methodology should deliver a packet from one node to all other network nodes using as few messages as possible. In this paper, we propose the Optimized Flooding Protocol (OFP), based on a variation of *The Covering Problem,* which is encountered in geometry, to minimize the unnecessary transmissions drastically and still be able to cover the whole region. OFP does not need *hello messages* and hence OFP saves a significant amount of wireless bandwidth and incurs lesser overhead. We present simulation results to show the efficiency of OFP. Moreover, OFP is scalable with respect to density; in fact OFP requires lesser number of transmissions at higher densities. OFP is also resilient to transmission errors.**

## I. INTRODUCTION

A "mobile ad hoc network" (MANET) is an autonomous system of mobile routers (and associated nodes) connected by wireless links--the union of which forms an arbitrary graph. The routers are free to move randomly and organize themselves arbitrarily; thus, the network's wireless topology may change rapidly and unpredictably. Such a network may operate in a standalone fashion, or may be connected to the larger Internet.

Flooding or Network wide broadcasting is the process in which one node sends a packet to all other nodes in the network. Many applications as well as various unicast routing protocols such as Dynamic Source Routing (DSR), Ad Hoc On Demand Distance Vector (AODV), Zone Routing Protocol (ZRP), and Location Aided Routing (LAR) use broadcasting or a derivation of it. The principal use of flooding in these protocols is for Location Discovery and for establishing routes.

A straightforward approach for broadcasting is *blind flooding*, in which each node will be required to rebroadcast the packet whenever it receives the packet for the first time. Blind flooding will generate many redundant transmissions, which may cause a more serious *broadcast storm problem* [4]. Given the expensive and limited nature of wireless resources such as bandwidth and battery power, minimizing the control message overhead for route discovery is a high priority in protocol design.

Recently, a number of research groups have proposed more efficient broadcasting techniques. Centralized broadcasting schemes are presented in [7, 8, 9]. Algorithms in [12-16, 18] utilize neighborhood information to reduce redundant messages.

This paper presents a new protocol to minimize the number of transmissions needed for broadcasting by doing selective forwarding, where only a few selected nodes in the network do the broadcasting. It is assumed that each mobile node knows its location. Various techniques like GPS [2], Time Difference of Arrival [25], Angle of Arrival [26] and Received Signal Strength Indicator [24] have been proposed to enable a node to discern its relative location. Recently, a range-free cost-effective solution [23] has been proposed for the same cause. To "select" the transmitting nodes, we extend the *Covering Problem* [1], which deals with covering a region completely using minimum number of circles.

The key advantages of our protocol are: a) With OFP the number of transmissions required decreases as the density of the network increases; b) OFP minimizes the number of unnecessary transmissions and outperforms other variations of flooding; c) In OFP, a node does not need to know locations/addresses of all its neighbors and hence OFP does not impose any bandwidth overhead such as *hello* messages; d) Behavior of OFP in large networks has been presented and it is shown that OFP performs well even in very large networks; e) OFP is able to reach a large fraction of nodes even when the nodes are moving at high speeds; f) OFP is robust to transmission errors as shown by our simulation results. Because of the above-mentioned advantages, OFP can also be used as an efficient broadcast protocol for Sensor Networks that operate in adverse conditions.

The rest of this paper is organized as follows: Section 2 discusses related work, Section 3 introduces *The Covering Problem* and a modification of *the Covering Problem*, Section 4 our approach for optimal flooding, Section 5 presents the simulation results of OFP and Section 6 concludes.

## II. RELATED WORK

Network-wide broadcast is an essential feature for ad hoc networks. The simplest method for broadcast service is flooding. Its advantages are its simplicity and reachability. However, for a single broadcast, flooding generates abundant retransmissions resulting in battery power and bandwidth waste. Also, the retransmissions of close nodes are likely to happen at the same time. As a result, flooding quickly leads to message collisions and channel contention. This is known as the broadcast storm problem [4].

The broadcast problem has been extensively studied for multi-hop networks. *Optimal* solutions to compute Minimum Connected Domination Set (MCDS) [9] were obtained for the case when each node knows the topology of the entire network (*centralized* broadcast). In particular, several solutions have been presented in which the broadcast *time complexity* is investigated in detail. The broadcast protocol introduced in [7] completes the broadcast of a message in $O(D\log^2 n)$ steps, where 'D' is the diameter of the network and 'n' is the number of nodes in the network. From the result proved in [8], this protocol is optimal for networks with constant diameter. For networks with a larger diameter, a protocol by Gaber *et al.* [9] completes the broadcast within $O(D+\log^5 n)$ time slots, and it is optimal for networks with $D \in \Omega(\log^5 n)$. These solutions are *deterministic* and guarantee a bounded delay on message delivery, but the requirement that each node must know the entire network topology is a strong condition, impractical to maintain in *ad hoc* mobile environments.

Several broadcast protocols that do not require the knowledge of the entire network topology have been proposed. In a counter-based scheme [4], a node does not retransmit if it overhears the same message from its neighbors for more than a prefixed number of times and in a distance-based scheme [4], a node discards its retransmission if it overhears a neighbor within a distance *threshold* retransmitting the same message.

Source Based Algorithm [14], Dominant Pruning [12], Multipoint Relaying [16], Ad Hoc Broadcast Protocol [15], Lightweight and Efficient Network-Wide Broadcast Protocol [18] utilize 2-hop neighbor knowledge to reduce number of transmissions.

A good classification and comparison of most of the proposed protocols is presented in [20]. It is also concluded that Scalable Broadcast Algorithm (SBA) [14] and Ad Hoc Broadcast Protocol (AHBP) [15] perform very well as the number of nodes in the network is increased. Both these techniques are based on two-hop neighbor knowledge.

The Scalable Broadcast Algorithm [14] requires that all nodes have knowledge of their neighbors within a two-hop radius. This neighbor knowledge coupled with the identity of the node from which a packet is received allows a receiving node to determine if it would reach additional nodes by re-broadcasting. Two-hop neighbor knowledge is achievable via periodic *hello messages*; each *hello messages* contains the node's identifier (IP address) and the list of known neighbors. After a node receives a *hello messages* from all its neighbors, it has two-hop topology information centered at itself.

AHBP [15] also requires that all nodes have knowledge of their neighbors within a two-hop radius. In AHBP, only nodes that are designated as a Broadcast Relay Gateway (BRG) within a broadcast packet header are allowed to rebroadcast the packet. BRGs are proactively chosen from each upstream sender, which is a BRG itself. A BRG selects set of 1-hop neighbors that most efficiently reach all nodes within the two-hop neighborhood as subsequent BRGs. Location Aided Broadcast [21] presents three location-aided broadcast protocols to improve communication overhead and the shortcomings of various protocols are also summarized.

In self-pruning methods [14, 19, 13], each node makes its local decision on forwarding status: forwarding or non-forwarding. Although these algorithms are based on similar ideas mentioned above, this similarity is not recognized or discussed in depth. Fair comparison of these algorithms is complicated by the lack of in-depth understanding of the effect of the underlying mechanisms.

The drawback of the above Neighbor Knowledge methods, which use local information to determine whether to rebroadcast, is their difficulty in mobile environments. Outdated 2-hop neighbor knowledge corrupts the determination of next-hop rebroadcasting nodes [20]. It should also be noted that conclusions in [20] were based on simulations on a network of area 3.5R X 3.5R, R being the range of the nodes. This implies that, in protocols based on 2-hop neighbor knowledge, nodes at the center of the network knows about 92% of the network topology; thus they can fairly approximate Minimum Connected Domination Set.

In Gossip-based routing [3], a node probabilistically forwards a packet so as to control the spreading of the packet through the network; the probability typically being around 0.65. Though, this simple mechanism reduces the number of redundant transmissions, it does not come close to the minimum transmissions achieved by centralized protocols and hence there is a lot of scope for further improvement.

In this paper we propose a new protocol, which needs minimal neighborhood information; neither the neighboring node addresses nor their locations are needed. This drastically reduces the effect of the mobility and also no *hello messages* are needed. Another property of OFP as illustrated through simulations is that the number of retransmitting nodes decreases as the number of nodes in the network increases. OFP is able to deliver a broadcast packet to large fraction of nodes even in highly mobile environments and in presence of transmission errors. OFP is resilient to transmission errors and radio propagation impairments.

## III. BACKGROUND

### A. *The Covering Problem*

*The Covering Problem* can be stated as follows:

"What is the minimum number of circles required to completely cover a given 2-dimensional space."

Kershner [1] showed that no arrangement of circles could cover the plane more efficiently than the hexagonal lattice arrangement shown in Fig. 1. Initially, the whole space is covered with regular hexagons, whose each side is R and then, circles are drawn to circumscribe them.

### B. *Modified-Covering problem*

Here, we state a modified version of *The Covering Problem* that finds its application in ad hoc networks. The solution we present here is to put forward the intuition behind our protocol and the solution is just for an ideal case scenario. A more practical solution is presented in section IV.

The modified version of t*he Covering Problem* can be stated as follows:

"What is the minimum number of circles of Radius R required to entirely cover a 2-dimensional space with the condition that the center of each circle being placed lies on the circumference of at least one other circle."

If the range of a mobile node is considered to be R, then the reason behind the condition that the center of a circle should lie on the center of another circle is that a Mobile Ad hoc node has to receive a message for it to retransmit the message. A possible solution for the *Modified-Covering Problem* is shown in Fig. 2. As done for covering problem, initially the whole region is covered with regular hexagons whose each side is R. Then, with each of the vertices as a center, circles of radius R are drawn.

The following properties of the vertices in Fig. 2 should be noted:

*Property-1:* Each vertex v is joined to three other vertices.

*Propery-2:* The lines joining these three vertices to vertex v make an angle of 120o (2π/3 radians) with each other.

*Propety-3:* Each vertex is at a distance of R from each of its neighboring vertices.

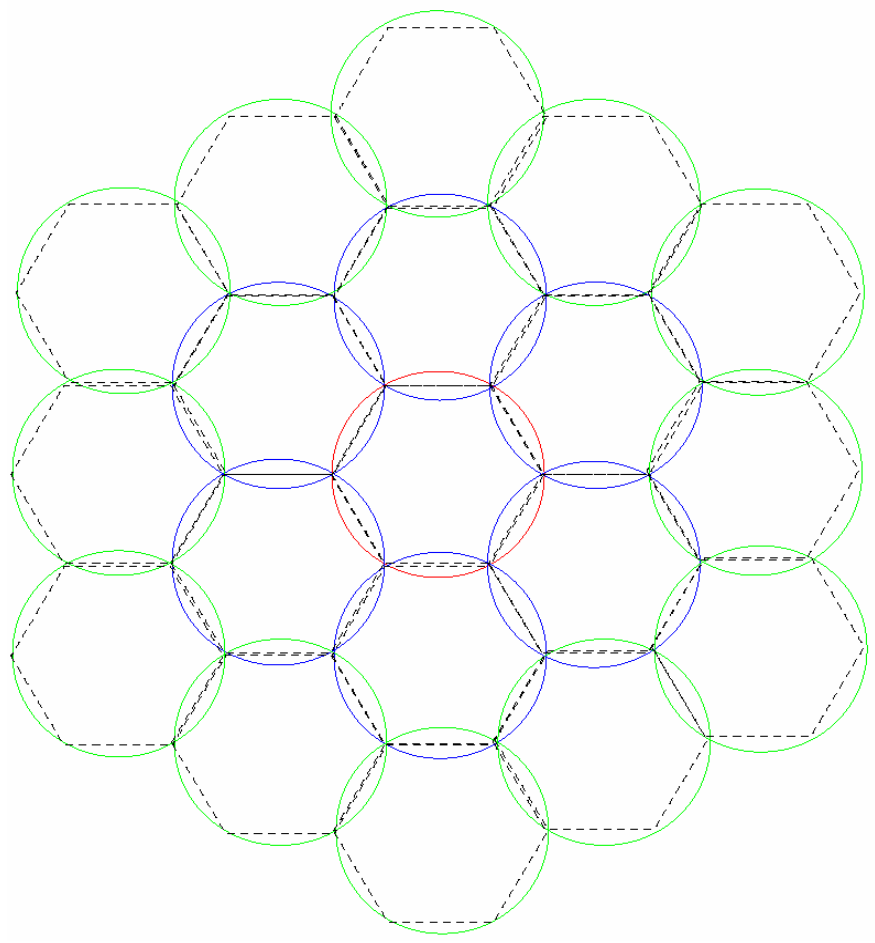

Fig. 1. Covering a plane with circles in an efficient way.

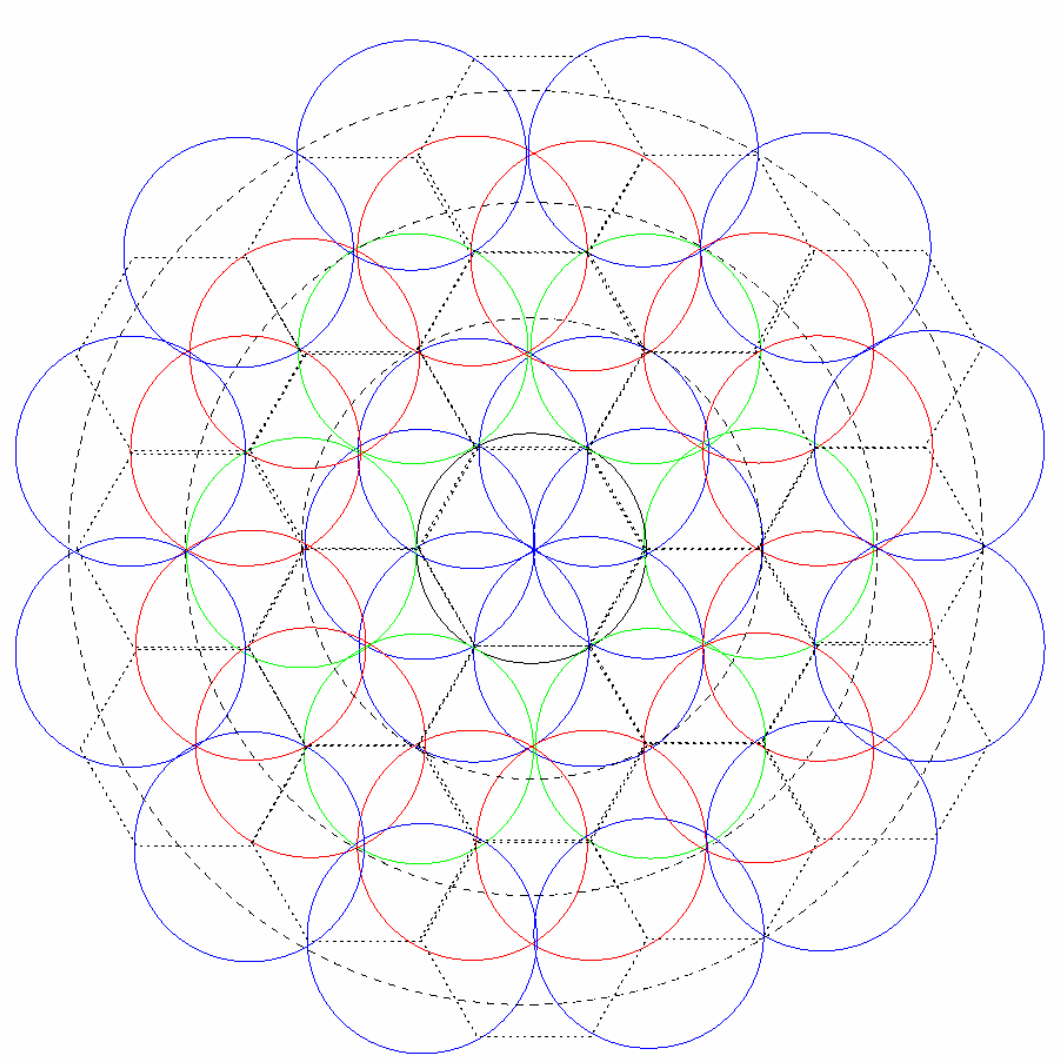

Fig. 2. Our Solution for the Modified-Covering Problem.

Thus, given a vertex *v* and one of its neighboring vertices, using the above properties, it is very easy to determine the other two neighboring vertices of vertex *v*.

The approach followed here to solve *the Modified-Covering problem* is for an ideal case scenario. We use the same approach to achieve broadcasting in a more general case, where there need not be any node at the optimal locations. In this case Fig-2 can get *skewed* a lot. For illustration, two such skewed figures are presented in Fig-5 in Section V. Even when the *skew* is very large, the number of transmissions required to cover the whole region remains very low.

Though we do not claim that the solution we presented for the *Modified-Covering problem* is the best, through simulations we show that our protocol implemented using this solution outperforms other broadcasting protocols. Also, we believe that the protocol can be easily adapted to any other geometric solutions of the *Modified-Covering Problem*.

## IV. Optimized Flooding Protocol (OFP)

In this section, we present the Optimized Flooding Protocol (OFP). Flooding achieves the goal of location discovery by letting all the nodes that receive the message, retransmit it again. The intuition behind our protocol is that in order to achieve the goal, there is no need for all nodes to transmit/retransmit the message. Instead, the goal can be achieved by allowing only a few strategically selected nodes to retransmit the message. The strategy to select such nodes is same as the strategy to solve the Modified Covering Problem presented in Section IIIB.

### A. Our Approach

Let S be the Source Mobile Node that sends the route request. As seen in Fig-2, after the first circle centered on the center of region (location of S), six more circles whose centers are located on circumference of the first circle are drawn. These can be considered as first stage retransmissions of the request. In the next stage again six more circles are drawn whose centers lie on the circumference of the circles drawn in the first stage. From now on using the properties 1, 2 and 3 presented in section IIIB, it is very easy to predict the centers of the circles to be drawn in the next stage.

In real life, though, it is seldom to have Mobile Nodes (MNs) located at the strategically selected locations. Thus, if the neighbor nodes are not in the optimal strategy locations, the coverage figure will get *skewed*; moreover, the *skew* effect may propagate. Our goal is to extend the Modified Covering Problem to meet this restriction. A simple solution is to select the nearest node to the point selected and that received the message to retransmit.

It should also be observed that a node could receive a message more than once – from different directions and from different nodes, each node specifying different optimal strategy location (because of the *skew*). This may cause two nodes very close to each other retransmit. We propose to avoid these transmissions by having a node keep track of its distance to the nearest node that has retransmitted the packet and to have a node retransmit only when its distance to the nearest transmitting node is greater than a threshold *Th*.

To elaborate for every broadcast packet, each node M stores the distance $d_m$ to the nearest node that has already transmitted the packet. A node does not retransmit, if $d_m$ for that broadcast message is less than a threshold *Th*. The choice of a right *threshold* will be the key for the success of the proposed algorithm. In section V.B, we study the performance of OFP with different threshold values and show that a *Th* value of 0.4*R is a good choice to ensure high delivery ratio while keeping the number of transmissions very low. R is the transmission range.

### B. The Algorithm

Each broadcast packet contains two location fields, $L_1$ and $L_2$ in its header. Whenever a node transmits a broadcast packet, it sets $L_1$ to the location of the node from which it received the packet and sets $L_2$ to its own location.

The Optimized Flooding Protocol is as follows:

The Source Node *S* sets both L1 and L2 to its location ($S_X$, $S_Y$, $S_Z$) and transmits the packet.

1. A node M, upon receiving a broadcast packet, first determines if the packet can be discarded. A packet can be discarded under any of the following conditions:
   - If the node has transmitted the packet earlier.
   - If $N_M$=1 i.e., the node from whom M received the packet is the only node in its range.
   - If a node which is very close has already transmitted this packet, i.e., if $d_n < Th$.
2. If the packet isn't discarded, M determines if it received the packet directly from the broadcast Source S.
   - If yes, M finds the nearest vertex V of a hexagon with ($S_X$, $S_Y$, $S_Z$) as its center and with ($S_X$+R, $S_Y$, $S_Z$) as one of its vertices. It computes its distance *l* from V and then delays the packet rebroadcast by a delay *d* given by $d = l/R$.
   - Else, if M hasn't received the packet directly from the source S, but from some other node K, then using properties 1, 2 and 3 mentioned in section 3 and with the nearest strategic location. The packet transmission is delayed by $d = l/20*R$.
3. After delay *d*, M again determines if it has received the same packet again and if the packet can be discarded (for the same reasons mentioned above). Thus, delaying enables a node to decide if it is the nearest node to the strategic location. M updates L1 to location of the node from which it received the packet and L2 to its location, sets $d_n$ to zero and transmits, if the packet cannot be discarded.

The *delaying* is used to make a node decide if it is the nearest node to the strategic location. Low delay values decrease the time needed to broadcast a message all over the network, while high delay values help reduce redundant transmissions in instances where two nodes are of about same distance from the strategic location. The delay function we used causes a packet to be delayed a maximum of 50ms per retransmission, though typically this value lies around 10ms. In dense networks, the delay values are much less than 10ms.

The computational complexity of OFP is negligible; when compared to flooding, the major additional computation is finding the node's distance to the nearest optimal point according to *the modified covering problem*, which can be easily computed using properties 1-3 mentioned in section III. The only bandwidth overhead due to OFP is because of addition of new header fields to carry location information of two nodes which is not significant.

## V. Experimental Results

We have developed a simulator to evaluate the performance of our protocol. We chose to compare our protocol with Ad Hoc Broadcast Protocol (AHBP) [15] as AHBP is one of the protocols (SBA [14] being the other) that approximates MCDS fairly [20]. A Mobile Ad Hoc Network of different physical areas and different shapes with different number of nodes were simulated. To be more specific, circular regions of radius varying from R to 10R and rectangular/square regions of size varying from 3R x 3Rm to 10R x 10R have been simulated, where R is the transmission range of each mobile node, which is 300m in all our simulations.

The nodes were uniformly distributed all over the region with the density varying from 4 nodes per R x R region to 100 MNs per R X R region. Every simulation is repeated until the 95% confidence intervals of all average results are within ±5%.

The simulations are aimed at studying the performance of OFP in networks of different sizes and densities. Initially, we simulated *the ideal case* where some node always exists at the strategically selected location. Then, we studied the effect of different threshold values on the performance of OFP. Then, we concentrated on the algorithm efficiency by studying the performance of OFP in static networks and also in highly mobile networks. We also present the effect of *hello message interval* on the performance of OFP. Lastly, we study the performance of OFP in networks where the coverage area of a node is not circular. The simulation results under each network study are presented in a subsection below.

### *A. Ideal Case Scenario*

We define *Ideal Case scenario* as a scenario in which some node exists exactly at each of the strategically selected locations.

The number of transmissions required to cover circular and rectangular regions in the ideal case scenario are observed and are as presented in Table-1(A) and Table-1(B). The number of transmissions required in the Ideal case present a lower bound on the number of transmissions required. As the density of the network increases the number of transmissions required approaches the lower bound.

### *B. Effect of Threshold Th*

The purpose of this study is to evaluate the effect of different threshold values on the performance of OFP. Figures 3 and 4 show the simulation results for threshold values of 0.35, 0.40 and 0.45. Apart from the number of transmissions in each case, the delivery ratio in percentage for each case is indicated at each data point. *Delivery Ratio* is the average number of nodes that receive the message to the total number of nodes in the network. Figure 3 is for a network size of 1800m X 1800m and Fig. 4 is for a network of size 1200m X 1200m.

Table –1(A) Number of transmissions required to cover a circular area in an Ideal Case

| Radius of Circular region | Number of transmissions |
|---|---|
| 2R | 12 |
| 3R | 24 |
| 4R | 42 |
| 5R | 60 |
| 6R | 90 |
| 7R | 126 |
| 8R | 168 |

Table-1(B) Number of transmissions requred to cover a rectangular area in an Ideal Case

| Size of the rectangular region | Number of Transmissions |
|---|---|
| 3R X 3R | 8 |
| 4R X 4R | 10 |
| 5R X 5R | 16 |
| 6R X 6R | 26 |
| 8R X 8R | 42 |
| 10R X 10R | 74 |
| 4R X 6R | 18 |
| 6R X 8R | 36 |
| 8R X 10R | 54 |

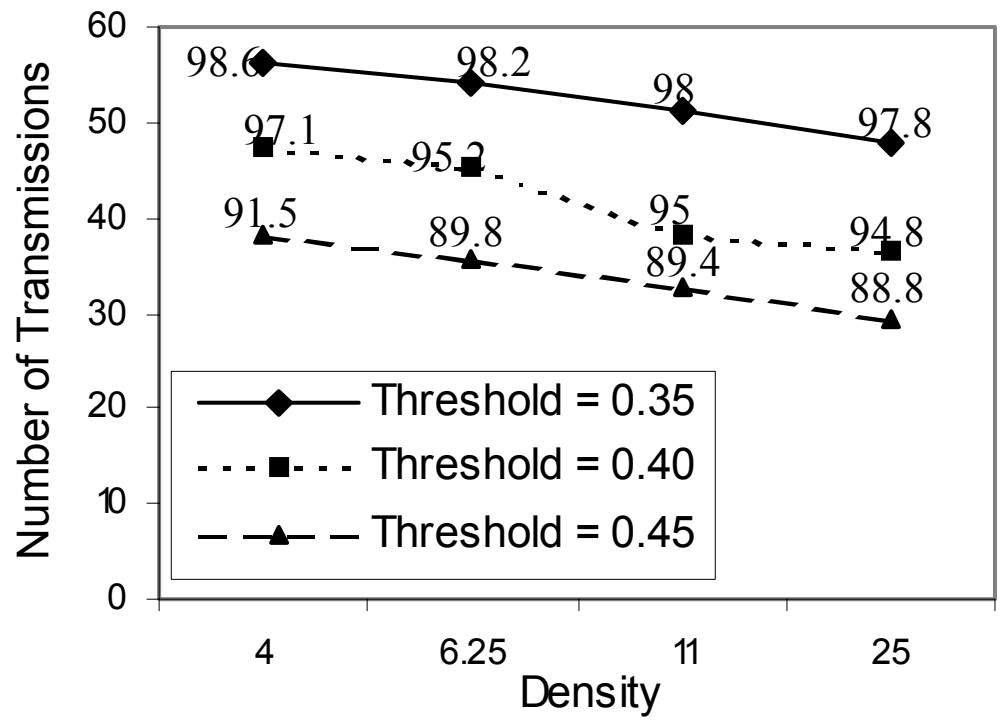

Fig. 3. Effect of Threshold on performance of OFP. Network size is 1800m X 1800m.

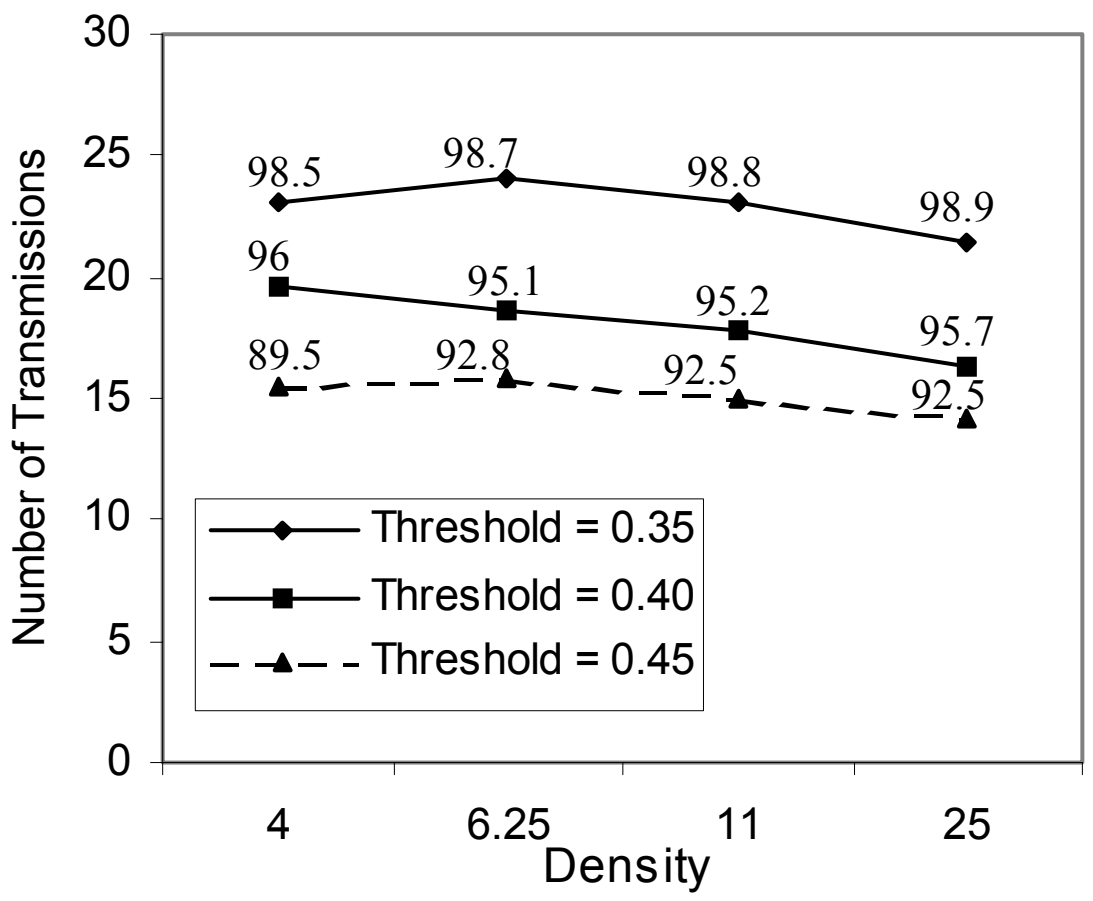

Fig. 4. Effect of Threshold on performance of OFP. Network size is 1200m X 1200m.

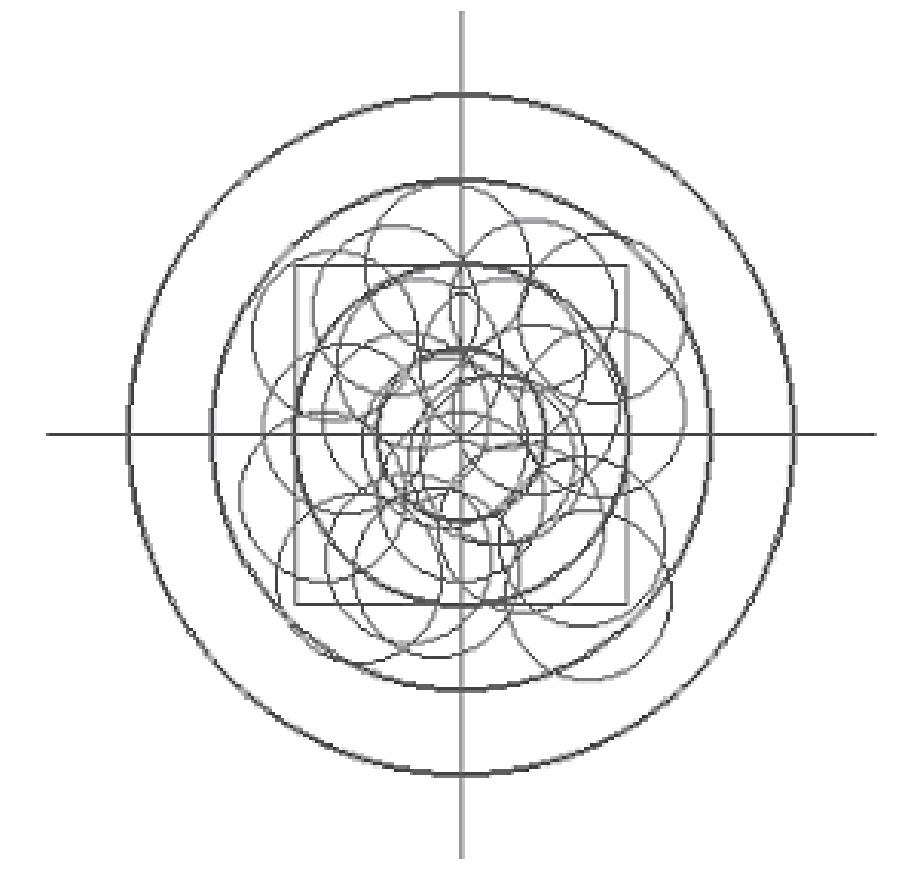

Fig. 5(a). An example skewed figure for 4R X 4R region with 64 nodes. Number of transmissions is 21.

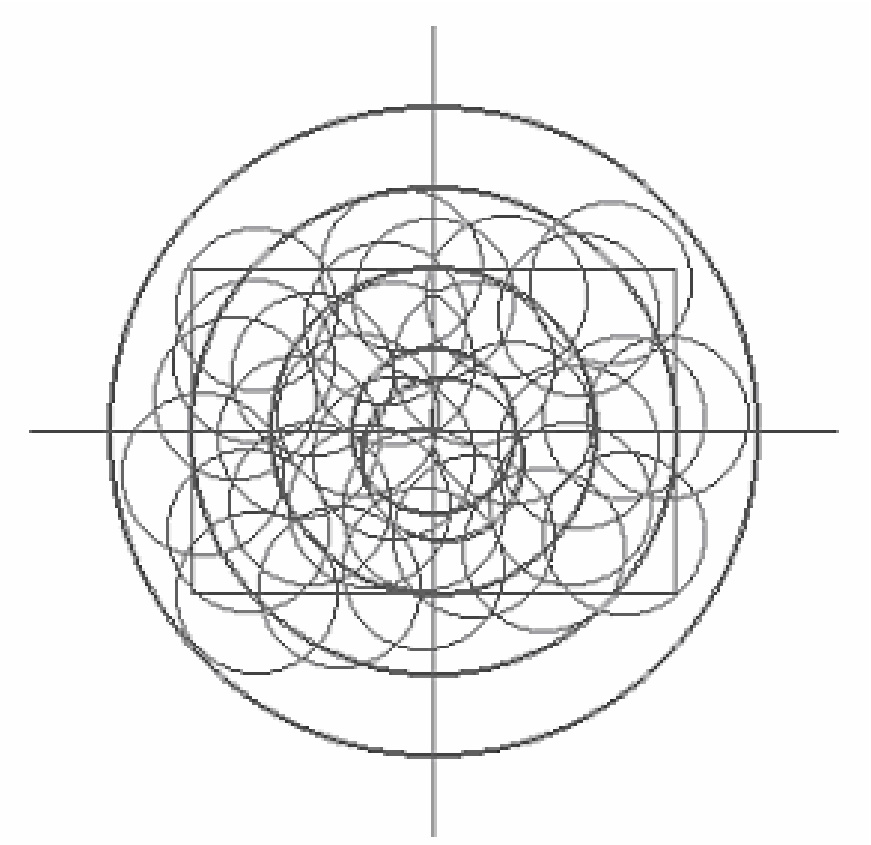

Fig. 5(b). An example skewed figure for 6R X 4R region with 96 nodes. Number of transmissions is 32.

For a threshold value of Th = 0.35, a delivery ratio of around 98% is achieved and for Th = 0.4, the delivery ratio is around 95%. But, for Th = 0.45, the delivery ratio falls to around 90%. This is understandable, because with the increase in threshold value, number of retransmitting nodes decrease.

For all further simulations, we use threshold value of Th = 0.4 and for each simulation case, we present the minimum and maximum delivery ratio, instead of presenting the delivery ratio for each for each data point.

### C. *OFP Efficiency*

The purpose of this study is to evaluate the performance of OFP in networks of different sizes and different densities. We include a "best-case" bound provided by the simulation results in *ideal case scenarios*. It is impossible for any algorithm to perform better than the performance in *ideal case scenario* and unlikely to perform worse than simple flooding. Thus, these two bounds provide a useful spectrum to gauge the performance of our protocol. For this study we varied the network size from 900m X 900m to 3000m X 3000m, while keeping the transmission radius of each node fixed to 300m. We also varied the density of the network from 4-nodes/R X R region to 100-nodes/R X R region.

First, fixing the density of the MNs in the region, we simulated the number of transmissions needed to cover a square/rectangular region completely. The coverage figure gets *skewed* a lot as in most of the cases no node exists at the strategic location. Fig. 4 shows two such cases – one for 4R X 4R and another for 6R X 4R regions, both with a density of 4 nodes per R X R region.

Figure 6 is a plot between the number of transmissions required to cover entire region for varying densities and for different areas of the region. Network areas up to 3000m X 3000m have been considered. Fig. 7 presents the results in a different perspective. It gives a plot between the number of transmissions and density of the network for different network sizes. It can be seen that the number of transmissions required decreases as the number of nodes (density) increases. The number of transmissions at a density of 100 is very near to the number needed in an *Ideal case*. The minimum delivery ratio achieved by OFP was 94.3% for the case with network size of 1800m X 2400m and with a density of 6.25. In all other cases, the delivery ratio was close to 95% with the maximum being 97.3%. The results show that the performance of OFP remains very efficient even in large networks; network size does not seem to affect the performance of OFP.

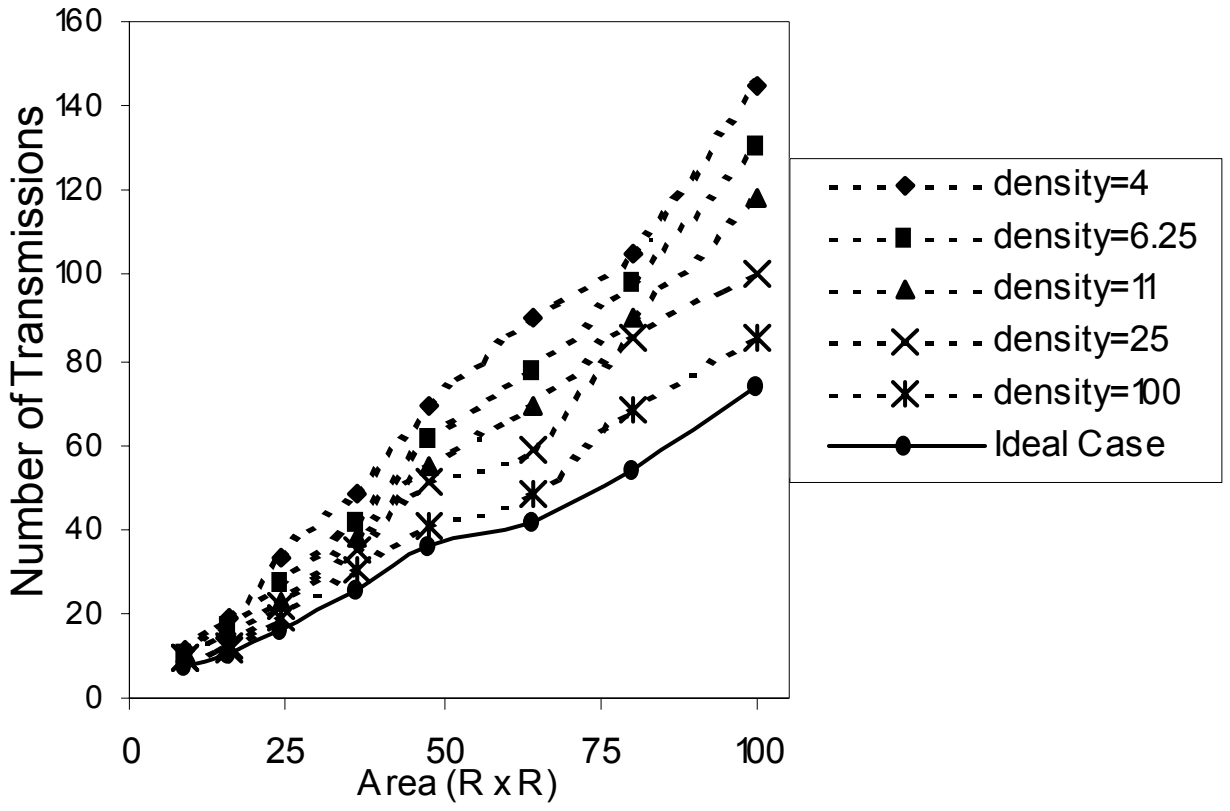

Fig. 6. Number of transmissions required to cover an entire region for different areas

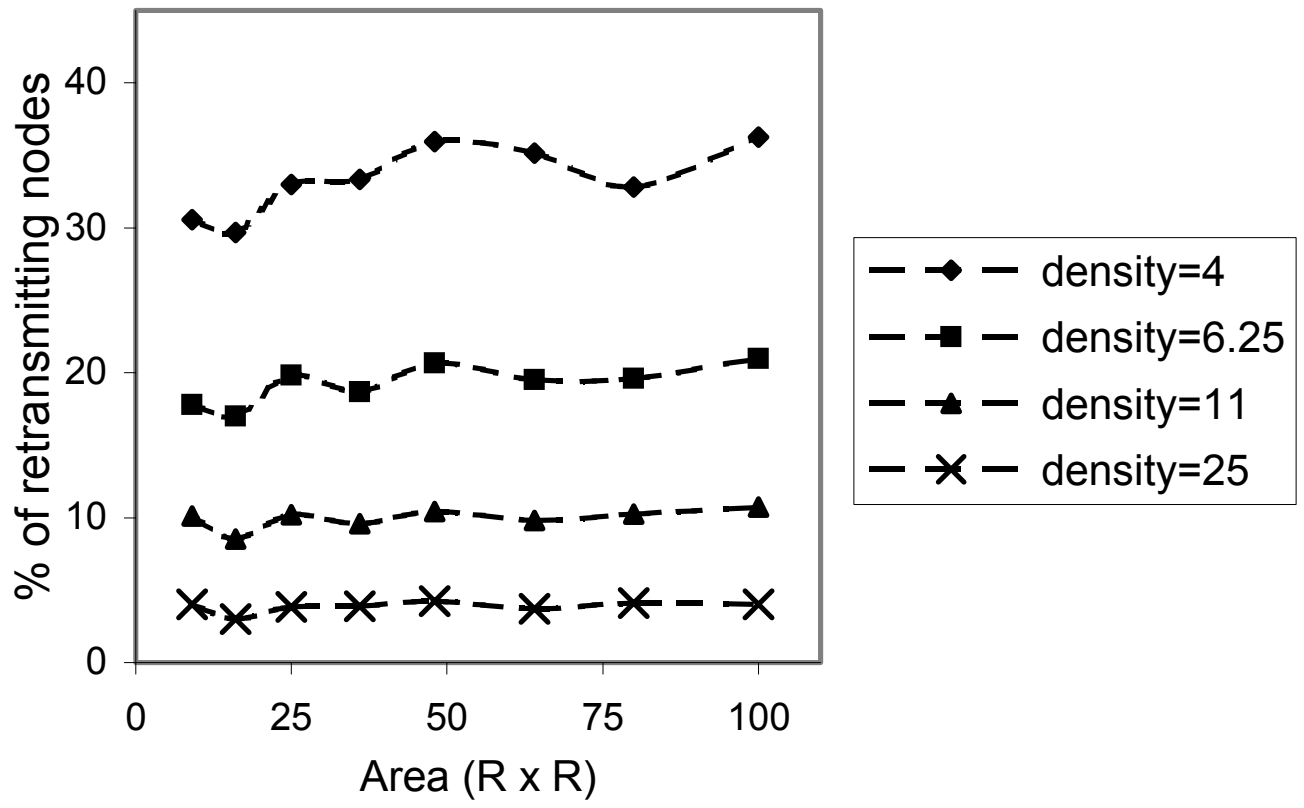

Fig. 8. Percentage of retransmitting nodes for different networks

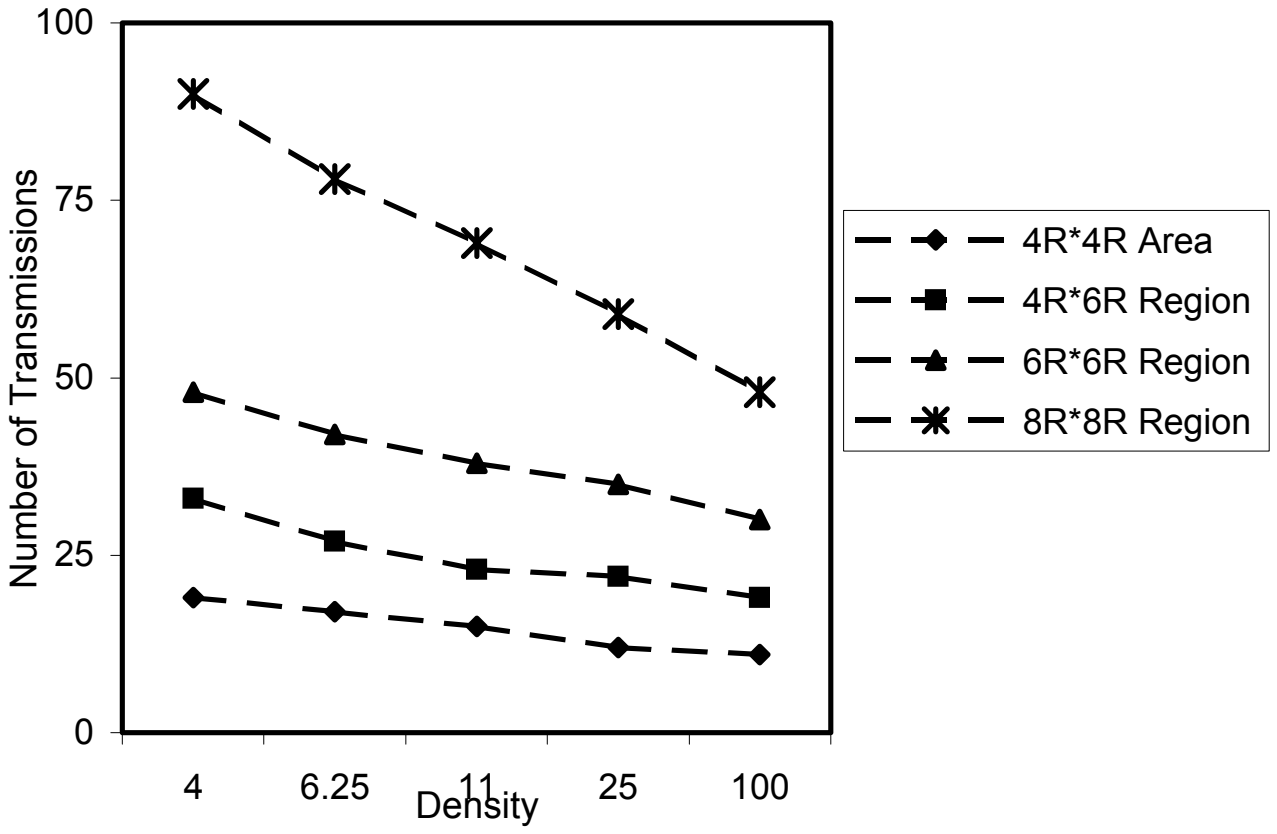

Fig. 7. Number of Transmissions for varying node densities and for different areas

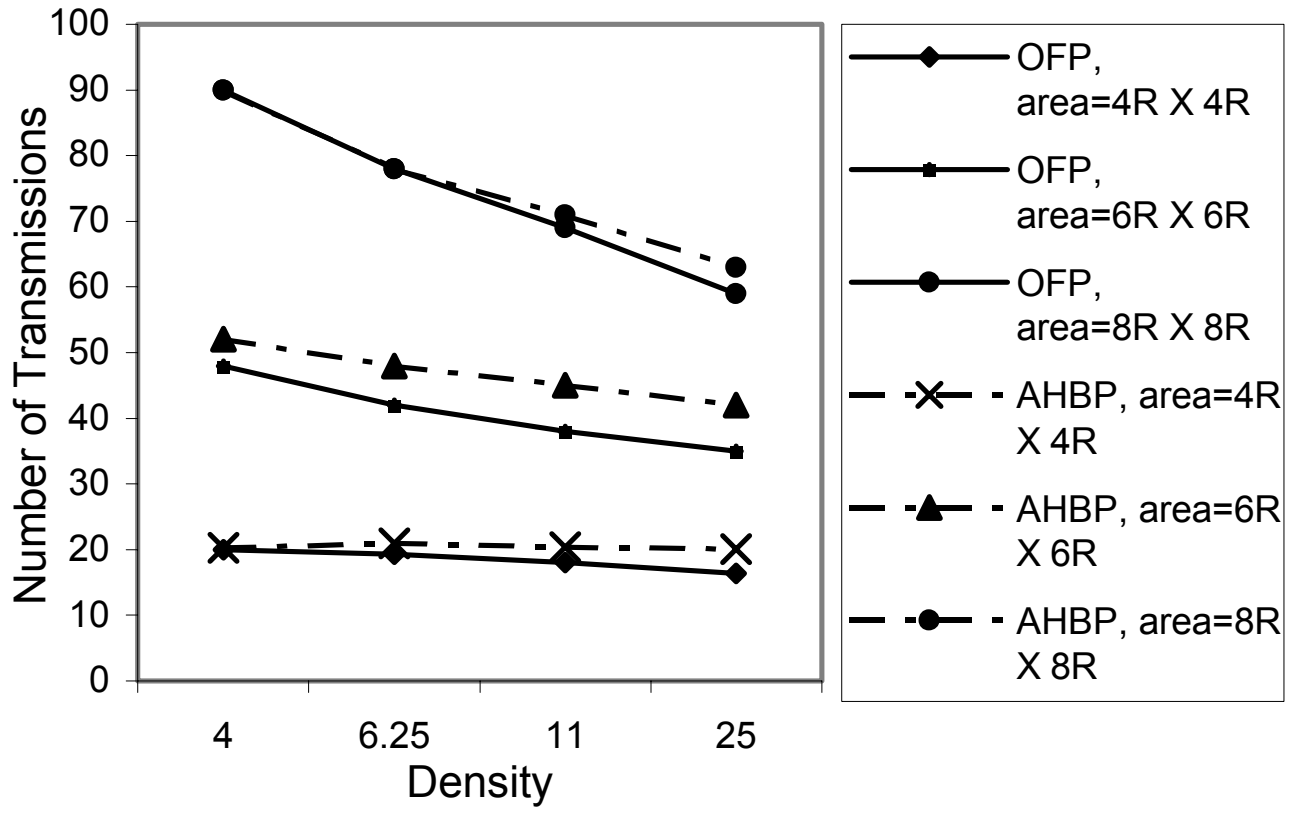

Fig. 9. Performance of OFP and AHBP in static networks

Figure 8 shows the percentage of nodes in the network retransmitting a broadcast message. The simulations were done in networks of sizes up to 3000m X 3000m with different node densities. For a given network density, the percentage of retransmissions remains almost a constant for all network sizes. This reflects that OFP performance is not hindered in large networks.

Next, we compare OFP with Ad Hoc Broadcast protocol (AHBP) [14]. Networks of 1200m X 1200m, 1800m X 1800m and 2400m X 2400m were considered. As shown by Fig. 9, the performance of both OFP and AHBP is very similar, though OFP performs slightly better than AHBP especially at high network densities. Here, we considered only static networks and in the next section, we present results for mobile networks where OFP clearly performs much better than AHBP.

### D. Mobile Networks

This section presents the simulation results of OFP and AHBP in mobile networks. We use the Random Walk mobility model [23] with zero pause time. The range of mean speeds of the nodes is varied from 1 to 20 meters per second. The upper bound corresponds to around 50 miles per hour, which we assume to be a realistic maximum speed of any mobile node.

Fig. 10 presents the effect of mobility on each of the protocols. The simulation is in a network 144 nodes and with the network size being 2400m X 2400m. The performance of OFP remains unaffected, as OFP algorithm uses minimal neighborhood information. But, the performance of AHBP rapidly deteriorates with increase in speed and its performance is also affected by the *hello interval*.

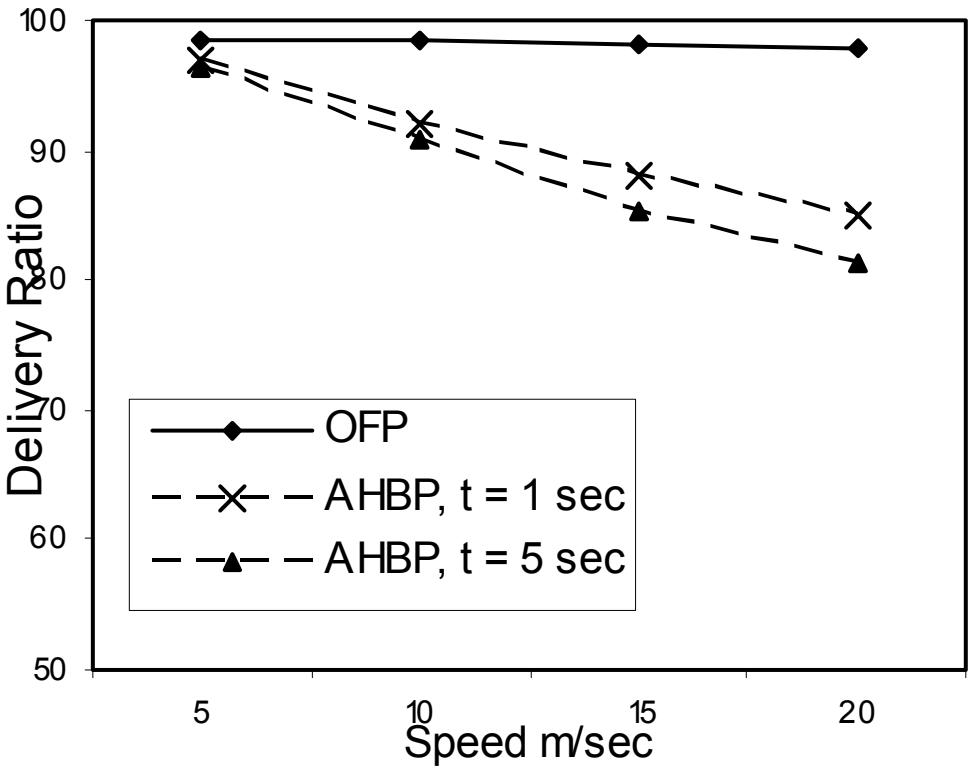

Fig. 10. Effect of Mobility on different protocols. Network size = 1800m X 1800m. Number of nodes =144.

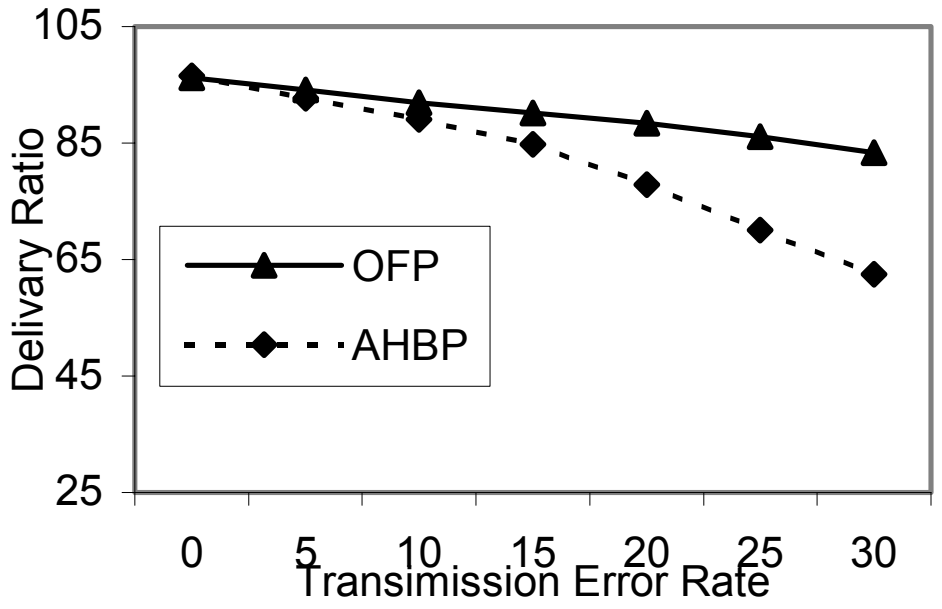

Fig. 11. Performance comparison of OFP and AHBP in presence Transmission errors. Network size = 1800m X 1800m. Number of nodes =144.

The two-hop neighbor knowledge based protocols use *hello messages* to gather the neighborhood information. With a hello interval of *t* seconds, the two-hop neighbor information (that is obtained through the hello messages of one-hop neighbors) would always be outdated by an average of *t* seconds. For instance, if $t$ = 10 seconds and a nodes speed is 36mph, then the node would have moved up to 100m before its information has been conveyed to one of its 2-hop neighbors. Also, once a node gets this information, it is not updated again till 10 sec. Thus, a node could have moved up to 200m before its information is updated at its neighbors. Also, the average time by which a node's information at 2-hop neighbor is out-dated is 15 seconds ($t + (0 + t)/2$), which corresponds to a displacement up to 150m. This shows the intensity of the effect mobility has on these protocols. Thus, the hello interval *t* should be very small for efficient performance of two-hop neighbor knowledge based protocols, which in turn means that the bandwidth overhead due to *hello messages* is very high.

### *E. Effect of Transmission Errors*

Wireless networks are characterized by losses due to transmission errors. We simulated the performance of OFP in networks with errors in transmission. Fig 11 compares the performance of OFP and AHBP in a network of size 1800m X 1800m. In the simulations, we placed 144 nodes randomly in the network. These simulations were for static networks. Transmission error rates up to 30% were simulated and we simulated Uniform transmission error model. It can be seen that the performance of OFP degrades gracefully with increase in transmission errors and OFP was able to achieve a delivery ratio of 84% even at a transmission error rate of 30%. At the same time, performance of AHBP degrades rapidly and the delivery ratio is less than 63% at an error rate of 30%.

The results show the robustness and resilience of OFP. This makes OFP a good choice for ad hoc networks that operate in adverse conditions. The high delivery ratio of OFP can be attributed to the fact that each node decides on its own whether to retransmit a packet or not and the decision is based on minimal neighborhood information brought by packets themselves. In presence of transmission errors, the closest node to the strategic location that has received the packet properly will retransmit. Also, one might expect that at a transmission error rate of 30%, on an average around 30% of the nodes would not be able to get the packet error free and the delivery ratio should be less than 70%. But, it should be noted that most of the nodes in the network receive a packet more than once and from different directions and hence, delivery ratio would be significantly better than 70%.

In case of AHBP, each retransmitting node recursively designates some of neighbors as Broadcast Relay Gateways (BRGs) and piggybacks the designated node addresses in the broadcast packet. Thus, if a designated BRG fails to receive the packet error-free, then no other node will be retransmitting instead of this node. Thus, the effect of transmission errors is much more profound on AHBP than OFP.

### *F. Effect of non-uniform Radio propagation*

In this section, we study the performance of OFP in wireless networks where wireless propagation is non-circular. We use the term *non-circularity* to mean that the range of a node might be different in each direction, the maximum being R, which is the range in an ideal case. Contours of the terrain and obstructions like large buildings contribute in creating such non-uniform radio propagation. We think this sort of study is necessary, especially as our protocol is an extension of the Modified Covering Problem solution developed for an ideal case.

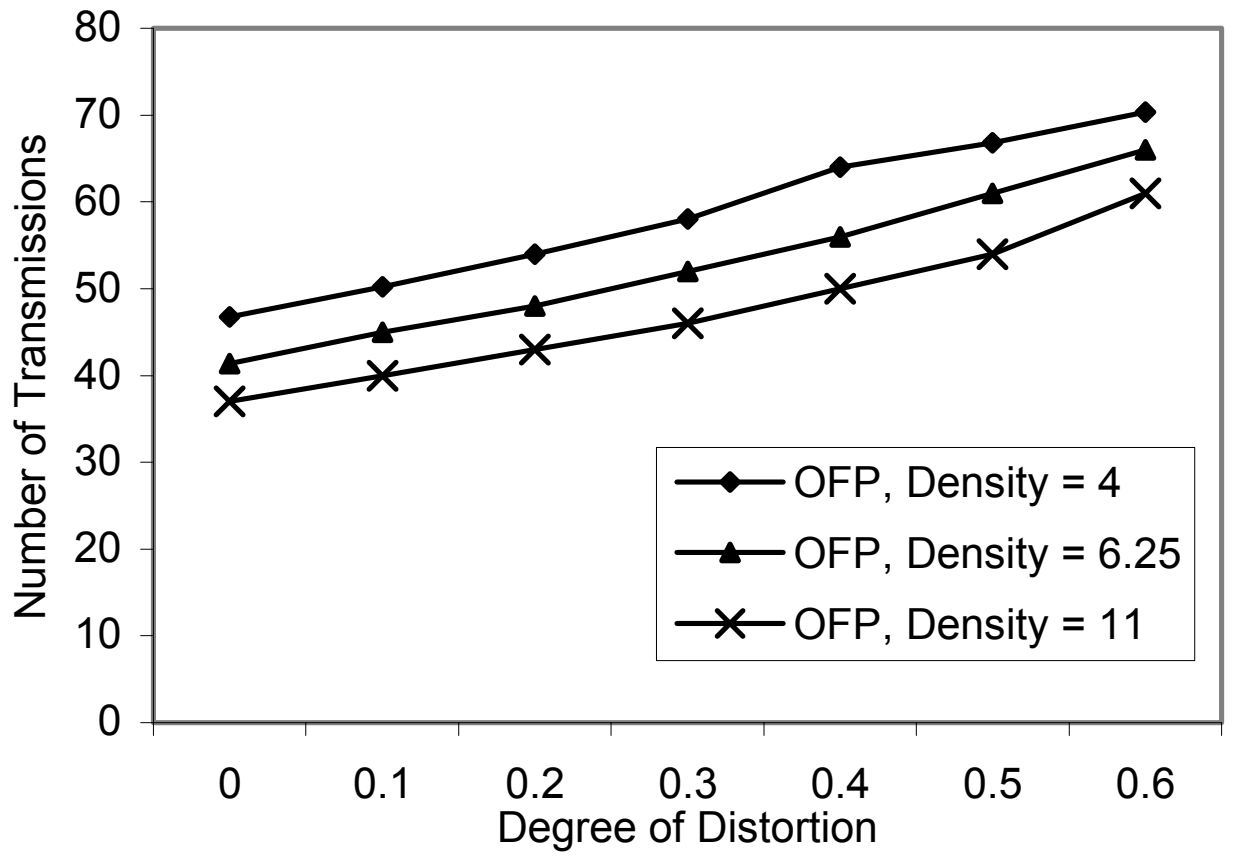

Fig. 12. Effect of non-uniform propagation on OFP. Network size is 1800m X 1800m

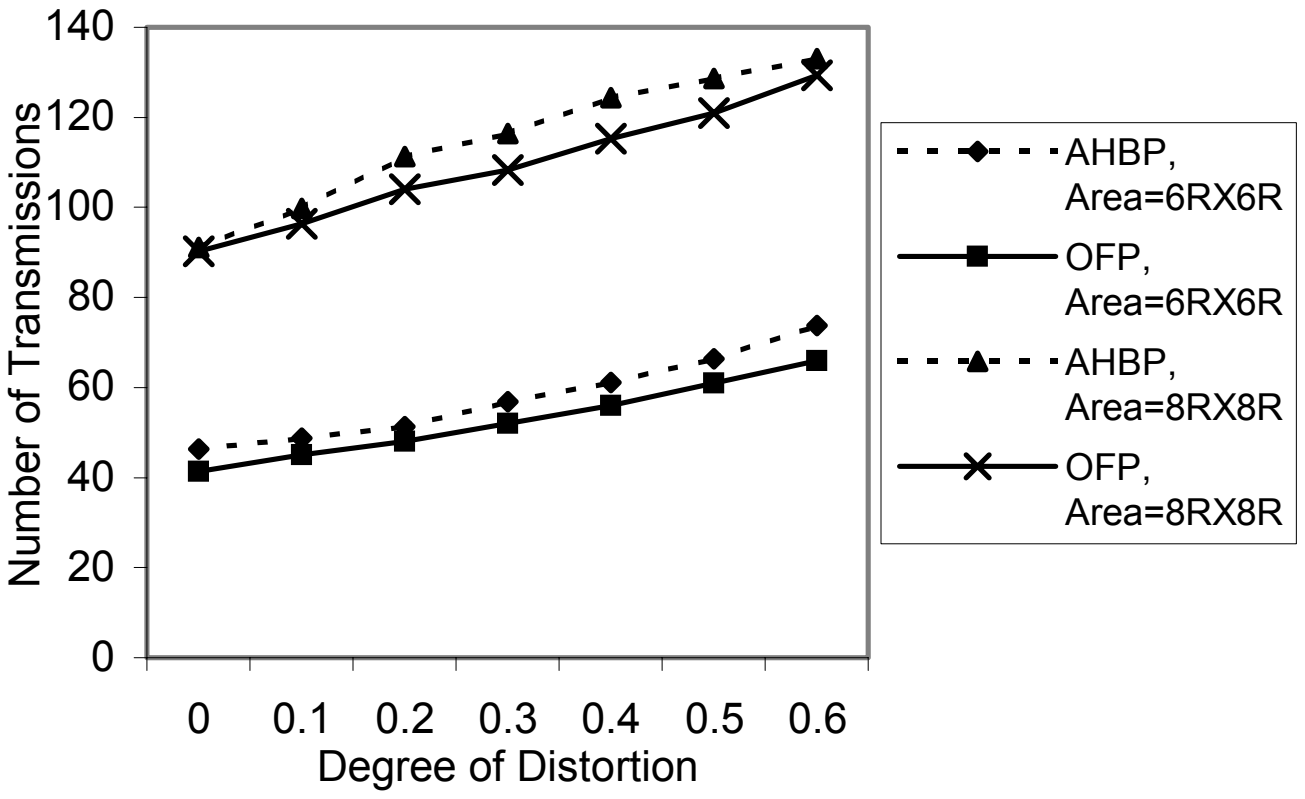

Fig. 13. Performance comparison of OFP and AHBP under non-uniform propagation. Network density = 6.25.

In the simulations, for each node, we generated the coverage area by setting the transmission range in different directions to a random value between [D*R, R], where D is the Degree of Distortion and R is the range of a mobile node in an ideal scenario. The simulations were for static networks.

The performance of OFP in case of non-circularity is presented in Fig. 12. Fig 12 is for a network area of 1800m X 1800m. It can be observed that the number of transmissions needed grow linearly with the degree of distortion. The delivery ratio in all cases was above 94% with the least being around 94.3%.

The performance comparison of OFP and AHBP is presented in Fig. 13. The figure is a plot between the number of transmissions and Degree of Distortion for network sizes of 1800m X 1800m and 2400m X 2400m and for a network density of 6.25. Performance of both the protocols is similar. In both protocols, the number of transmissions increases almost linearly with respect to the Degree of Distortion. The effect of mobility is not considered in these simulations.

The purpose of the study was to see the performance of OFP in networks with non-uniform transmission ranges. As shown by figures 12 and 13, OFP's performance remains efficient even under such conditions. This can be attributed to fact that in OFP, the decision if a node retransmits or not is made locally at each node that receives the packet. Thus, even if a node very close to the strategic location does not get the packet, the reachability is not affected as some other node that received the packet retransmits.

## VI. CONCLUSION

Building efficient broadcast protocols for ad hoc networks is challenging due to the dynamic nature of the nodes. In this paper we proposed Optimized Flooding Protocol (OFP), a novel protocol for broadcasting. The protocol is based on a variation of the Covering Problem. OFP is performed in an asynchronous and distributed manner by each node in the network.

OFP has a number of advantages over other approaches considered in the literature. The best feature of OFP is that a node needs only minimal local information to make a propagation decision and hence, OFP does not impose any bandwidth overhead in terms of *hello messages*. The efficiency of OFP remains very high even in large networks and OFP scales with density. Its efficiency in mobile networks and its robustness even in presence of transmission errors make it an ideal choice for MANETs and sensor networks.